\documentclass{LMCS}
\usepackage{amsmath,amssymb}

\usepackage{enumerate}
\usepackage{hyperref}

\newcommand{\SetN}{\ensuremath{\mathbb{N}}}

\newcommand{\bigoh}[1]{\ensuremath{O(#1)}}

\newcommand{\condset}[2]{\ensuremath{\left\{#1\enspace\left|\enspace#2\right.\right\}}}

\newcommand{\rd}{\ensuremath{\;\leftarrow\;}}

\newcommand{\tp}{\operatorname{tp}}
\newcommand{\cl}{\operatorname{cl}}
\newcommand{\ucl}{\operatorname{ucl}}

\newcommand{\norm}[1]{\ensuremath{||#1||_{\infty}}}

\numberwithin{equation}{section}

\def\doi{5 (1:4) 2009}
\lmcsheading%
{\doi}
{1--21}
{}
{}
{Dec.~20, 2006}
{Feb.~27, 2009}
{}   

\begin{document}
\title{The Complexity of Datalog on Linear Orders}
\author[M.~Grohe]{Martin Grohe\rsuper a}   
\address{{\lsuper a}Institut f\"ur Informatik, Humboldt-Universit\"at zu Berlin, Unter den
 Linden 6, 10099 Berlin, Germany}       
\email{grohe@informatik.hu-berlin.de}  

\author[G.~Schwandtner]{Goetz Schwandtner\rsuper b}      
\address{{\lsuper b}Institut f\"ur Informatik, Johannes Gutenberg-Universit\"at,
 55099 Mainz, Germany}  
\email{schwandtner@uni-mainz.de}  


\titlecomment{}


\begin{abstract}
  \noindent We study the program complexity of datalog on both finite and
  infinite linear orders. Our main result states that on all linear
  orders with at least two elements, the nonemptiness problem for
  datalog is EXPTIME-complete. While containment of the nonemptiness
  problem in EXPTIME is known for finite linear orders and actually
  for arbitrary finite structures, it is not obvious for infinite
  linear orders. It sharply contrasts the situation on other infinite
  structures; for example, the datalog nonemptiness problem on an infinite
  successor structure is undecidable. We extend
  our upper bound results to
  infinite linear orders with constants.
 
  As an application, we show that the datalog nonemptiness problem on
  Allen's interval algebra is EXPTIME-complete.
\end{abstract}

\maketitle 
\section{Introduction}
Datalog is the language of logic programming without
function symbols. Datalog has been extensively studied in database theory
(see, e.g.,
\cite{abihulvia95,chavar92,cosgaikanvar88,gaimaisgavar93,gotkoc04,kan90,kolvar95,gottlob03complexity}).  In
particular, the complexity of evaluating datalog queries has been determined:
The data complexity is PTIME-complete, and the program complexity (also known
as expression complexity) and combined
complexity are both EXPTIME-complete \cite{dantsin97complexity,var82}. 

While previous work on datalog was concerned with datalog over finite
structures, in this paper we are mainly interested in infinite structures.
Infinite structures occur naturally in spatial and temporal reasoning (and
spatial and temporal databases). In temporal reasoning, time is usually
modelled as an infinite linear order, sometimes discrete and sometimes dense.
This motivates our study of datalog on infinite linear orders. Let us remark
that our results also apply to an interval based temporal reasoning, carried
out, for example in Allen's interval algebra~\cite{allen82maintaining} (see
Sec.~\ref{sec:allall}).

When studying the complexity of datalog on infinite structures, we
consider the structure as fixed, that is, we are interested in program
complexity. Note that the result of a datalog query on an infinite
structure may be infinite, thus we cannot hope to compute the full
query result in finite time. A reasonable version of the query
evaluation problem that avoids this problem is the \emph{datalog tuple
  problem}, which asks whether a given tuple of elements is in the
result. However, even for the tuple problem there is the technical
issue of how to represent the elements of the tuple and how to
represent the structure itself. The simpler \emph{datalog nonemptiness
  problem}
asks if the result of a query is empty. It is
well known that on finite structures, the nonemptiness problem is in
EXP\-TIME, and as long as the structures have two elements that can be
distinguished by a datalog program, it is EXPTIME-complete (see
Sec.~\ref{sec:lowbnd}). It is easy to see that there are infinite
structures where the nonemptiness problem is undecidable. An example is the
structure with one infinite successor relation (see
Sec.~\ref{sec:lowbnd}).

Our first main result (Theorem~\ref{thm:discrtime}) states that on all linear
orders (finite or infinite), the datalog nonemptiness problem is decidable in
EXPTIME; for all linear orders with at least two elements it is
EXPTIME-complete.

A problem that has received considerable attention in the datalog literature
is the boundedness problem (see, e.g.,
\cite{abi89,gaimaisgavar93,hillebrand95undecidable}). A datalog program $\Pi$
is \emph{uniformly bounded on a class $C$ of structures} if there is a number
$b=b(\Pi,C)$ such that for all structures $\mathcal A$ in $C$, the computation
of $\Pi$ on $\mathcal A$ reaches a fixed point in at most $b$ steps. The
boundedness problem asks if a given program is uniformly bounded on the class
of all finite structures; it was shown to be undecidable in
\cite{gaimaisgavar93}.

Our second main result (Theorem~\ref{theo:ucl}) states that every datalog
program is uniformly bounded on the class of all linear orders. This also
leads to the decidability of the datalog tuple problem on linear
orders, provided the linear order satisfies certain effectivity conditions.

Technically, both results are based on an analysis of the
\emph{distance types} of tuples processed in the evaluation of a
datalog program. Types are a tool from model theory; the type of a
tuple of elements records ``definable'' information about this tuple.
Our distance types record information about the relative order of and
the pairwise distances between elements of a tuple. The crucial
technical fact underlying the results is that the whole computation of
a datalog program on a linear order can be described in terms of a
finite number of distance types that is bounded in terms of the
program (independently of the structure).

In the last section of this paper, we show how to incorporate
constants into the distance type concept to transfer our results to
datalog over linear orderings with a finite number of constants, which
may occur in the datalog programs in question.

As related results, let us mention recent results on the complexity of
constraint satisfaction problems on infinite structures
\cite{boddal06,bodnes06,krokhin04constraint}. With some handwaving, the
datalog nonemptiness problem may be viewed as a ``recursive version'' of
constraint satisfaction problems.\footnote{Our actual starting point was an
  attempt to understand the complexity of constraint logic programming, which
  combines logic programming and constraint satisfaction. It turned out,
  however, that this complexity is dominated by the complexity of the ``logic
  programming'' part, which then led to our interest in datalog.}

\section{Preliminaries}
\label{sec:def}

\subsection{Datalog}
An \emph{atom} is an expression of the form $P(x_1,\ldots,x_k)$, where $P$ is
$k$-ary relation symbol and $x_1,\ldots,x_k$ are variables. We admit $0$-ary relation
symbols.\footnote{A $0$-ary relation either is empty, or it consists of the
  empty tuple $()$.} In the following,
we abbreviate tuples $(x_1,\ldots,x_k)$ by $\bar x$.
A \emph{datalog rule} is an expression $\rho$ of the form
\[
P\bar{x}\rd Q_1\bar{y}_1,\dotsc,Q_{m}\bar{y}_{m},
\]
where $P\bar{x}$, $Q_1\bar{y}_1,\ldots,Q_{m}\bar{y}_{m}$ are
atoms. The tuples of variables $\bar x,\bar y_1,\ldots,\bar y_m$
need not be disjoint, and variables may be occur several times in each
tuple. Furthermore, the variables in $\bar x$ are not required to be
among those in $\bar y_1,\ldots,\bar y_m$. 
The atom $P\bar{x}$ is the \emph{head} of the rule and
$Q_1\bar{y}_1,\ldots,Q_{m}\bar{y}_{m}$ is the \emph{body}.  A
\emph{datalog program} is a set of datalog rules. Relation symbols occurring in the
head of a rule of a datalog program $\Pi$ are called \emph{intensional relation
  symbols} or \emph{IDBs}; all other relation symbols are called
\emph{extensional relation symbols} or \emph{EDBs}.

Datalog programs are interpreted over relational structures. A
\emph{vocabulary} is a finite set $\tau$ of relation symbols, each with a
fixed \emph{arity}. A structure $\mathcal A$ of vocabulary $\tau$ consists of
a (finite or infinite) set $A$ and a $k$-ary relation $R^{\mathcal A}$ for
every $k$-ary relation $R\in\tau$. We say that a datalog program $\Pi$ is
\emph{over} a structure $\mathcal A$ if the vocabulary of $\mathcal A$
contains all EDBs of $\Pi$ and none of the IDBs. $\Pi$ is a datalog program
\emph{over} a class $C$ of structures if $\Pi$ is a program over all $\mathcal
A\in C$.

Let $\Pi$ be a Datalog program over a structure $\mathcal A$. The
\emph{computation of $\Pi$ over $\mathcal A$} is carried out in
stages, in which the interpretation of the IDBs is computed; the
interpretation of the EDBs is given by $\mathcal A$ and remains fixed.
Initially, all IDBs are interpreted by the empty set. In each stage, a
rule $\rho$ of $\Pi$ is applied, and some tuples of elements of $A$
are added to the interpretation of the IDB occurring in in the head of
rule $\rho$. Formally, for every $k$-ary IDB $R$ we define a sequence
$(R_{i}^{\Pi,\mathcal A})_{i\ge 0}$ of $k$-ary relations on the
universe $A$ of $\mathcal A$. We let $R_0^{\Pi,\mathcal A}=\emptyset$
for all IDBs $R$. Suppose now we have defined $R_{i-1}^{\Pi,\mathcal
  A}$ for all IDBs $R$. In stage $i$, we choose a rule $\rho$, say,
$P\bar{x}\rd Q_1\bar{y}_1,\dotsc,Q_{m}\bar{y}_{m}$.  An
\emph{instantiation} of $\rho$ at stage $i$ consists of tuples $\bar
a,\bar b_1,\ldots,\bar b_m$ of elements of $A$ matching the lengths of
the variable tuples $\bar x,\bar y_1,\ldots,\bar y_m$, such that
\begin{enumerate}[$\bullet$]
\item If two variables of the rule are equal, then the corresponding elements
  of the tuples are equal as well. For example, if
  $x_r=y_{st}$ then $a_r=b_{st}$.
\item For $1\le r\le m$: If $Q_r$ is an EDB, then $\bar b_r\in Q_r^{\mathcal
    A}$. If $Q_r$ is an IDB, then $\bar b_r\in Q_{r\,(i-1)}^{\Pi,\mathcal A}$.
\end{enumerate}
We let 
\[
P_i^{\Pi,\mathcal A}=P_{i-1}^{\Pi,\mathcal A}\cup\{\bar a\mid\parbox[t]{8cm}{there exist tuples $\bar
  b_1,\ldots,\bar b_m$ such that $\bar a,\bar
b_1,\ldots,\bar b_m$ is an instantiation of rule $\rho$ at stage $i\}.$}
\]
For all IDBs $R\neq P$, we let $R_i^{\Pi,\mathcal A}=R_{i-1}^{\Pi,\mathcal
  A}$. To turn this into a well-defined deterministic process, we cycle
through the rules $\rho$ of $\Pi$ in some fixed order. It can be shown that the
  result of the computation does not depend on this order. (It will be convenient
later to apply only one rule at each stage, that is why we set up the computation
this way.)
  
Note that for all IDBs $R$ and for all $i\ge 0$ we have $R_i^{\Pi,\mathcal
  A}\subseteq R_{i+1}^{\Pi,\mathcal A}$. The process either reaches a fixed point
after finitely many stages, that is, there is an $i_0$ such that
$R_{i_0}^{\Pi,\mathcal A}=R_{i}^{\Pi,\mathcal A}$ for all $i\ge i_0$, or it
continues forever (recall that $\mathcal A$ may be infinite). In both cases,
we let $R_{\infty}^{\Pi,\mathcal A}=\bigcup_{i\ge 0}R_i^{\Pi,\mathcal A}$.
Then the $R_{\infty}^{\Pi,\mathcal A}$ form a fixed point of the computation,
that is, further applications of the rules do not increase the relations. This
is obvious if a fixed-point is reached in finitely many stages, but also easy
to see if not. The result of the computation is the interpretation of the IDBs
  in this fixed point.

We usually write $R_i^{\Pi}$ and $R_{\infty}^{\Pi}$ instead of
$R_{i}^{\Pi,\mathcal A}$ and $R_{\infty}^{\Pi,\mathcal A}$ if $\mathcal A$ is
clear from the context.  For an easier reference, we define the following set
of parameters for a datalog program $\Pi$: By $m_L$ we denote the maximal IDB
arity (i.e.  variables on the left hand side, head part of program rules), by
$m_R$ the maximal number of different variables occurring in a rule. By $n_R$
we denote the number of rules of $\Pi$ and by $n_I$ the number of IDB
symbols, by $m_I$ the maximal number of IDB occurrences in a rule body.
All these parameters are bounded from above by the length $n:=|\Pi|$ of $\Pi$
in some standard encoding.

For a more detailed introduction to datalog, we refer the reader to
\cite{abihulvia95}. 

\subsection{Linear orders}
A \emph{linearly ordered set} is a structure $\mathcal A=(A,<^{\mathcal A})$ of
vocabulary $\{<\}$, where the binary relation $<^{\mathcal A}$ is a linear order of the
universe $A$. For brevity, we refer to linearly ordered sets just as
\emph{linear orders}. Moreover, we usually omit the superscript in
$<^{\mathcal A}$ and use the symbol $<$ to denote both the relation
$<^{\mathcal A}$ and the relation symbol $<$. We write $a\le b$ instead of
($a<b$ or $a=b$). The \emph{distance} $d(a,b)$ between
two elements $a<b\in A$ is the maximum $d\ge0$ such that there are
elements $c_0,\ldots,c_d\in A$ with $a= c_0<c_1<\cdots<c_d=b$ if this
maximum exists, and $\infty$ otherwise. The linear order $(A,<)$ is \emph{dense without
  endpoints}, if for all $a\in A$ there are $b,c\in A$ such that
$b<a<c$, and for all $a,b\in A$ with $a<b$ there is a $c\in A$ such
that $a<c<b$.

We consider linear orders in the strict sense, that is, a linear order is
always antireflexive. For orders in the sense of ``less-than-or-equal-to'', the
datalog nonemptiness problem is trivial,\footnote{Actually, the problem is
  still PTIME-complete; it is equivalent to the datalog nonemptiness problem
  over a structure with one element, which is equivalent to the satisfiability
  problem for propositional Horn clauses.} because we can always satisfy all
atoms by interpreting all variables by the same element of the universe. 

\subsection{Algorithmic problems}\label{sec:problems}
We shall study the complexity of the following two decision problems for
fixed structures $\mathcal A$:

\medskip\noindent
\textbf{\itshape Datalog nonemptiness problem over $\mathcal A$}
\begin{quote}
  \begin{description}
  \item[Instance] Datalog program $\Pi$ over
    $\mathcal{A}$, IDB $P$ of $\Pi$.
  \item[Question] Is $P_\infty^{\Pi,\mathcal A}\ne\emptyset$?
  \end{description}
\end{quote}

\medskip\noindent
\textbf{\itshape Datalog tuple problem over $\mathcal A$}
\begin{quote}
  \begin{description}
  \item[Instance] Datalog program $\Pi$ over
    $\mathcal{A}$, $k$-ary IDB $P$ of $\Pi$, $k$-tuple $\bar a$ of elements of
    $\mathcal A$ (for some $k\ge 1$).
  \item[Question] Is $\bar a\in P_\infty^{\Pi,\mathcal A}$?
  \end{description}
\end{quote}
For an infinite structure (with finite vocabulary and finite EDB
and IDB arities, but infinite universe), the tuple problem bears some
difficulties with regards to the representation of the input tuple and
the accessibility of the structure. To deal with the first difficulty,
whenever we consider the tuple problem we assume that the universe of
the structure $\mathcal A$ is a decidable set of strings over some
finite alphabet.  Furthermore, for linear orders $\mathcal
A=(A,<^{\mathcal A})$ we assume that it is decidable whether for
elements $a,b\in A$ and a nonnegative integer $k$ there exist
$a_1,\dotsc,a_k\in A$ with $a<^{\mathcal{A}}a_1 <^{\mathcal{A}}\dotsb
<^{\mathcal{A}}a_k<^{\mathcal{A}}b$. Note that if this is undecidable,
then the datalog tuple problem over $\mathcal A$ is also undecidable.
Thus our assumption is just a restriction to the interesting cases of
the problem.  

Let us emphasise that these effectivity assumptions on the
representation are only required when we study the datalog tuple
problem. For the nonemptiness problem, we do not need to make any
assumptions on the representation or decidability of $\mathcal A$
whatsoever.

\section{Datalog on Allen's interval algebra}
\label{sec:allall}

Allen's interval algebra, introduced in \cite{allen82maintaining}, is
an algebra of relations over open intervals on the real line. These
interval relations are built as unions from the 13 basic relations
describing the pairwise relative end points of two intervals
$(x^-,x^+)$ and $(y^-,y^+)$ as Table \ref{tab:allensalg} (taken from
\cite{krokhin04constraint}). The algebra of these $2^{13}$ relations is equipped with the
operations \textit{converse} (denoted by $\cdot^{-1}$),
\textit{intersection} $\cap$ and \textit{composition} $\circ$. 

The complexity of constraint satisfaction problems over Allen's
interval algebra and variants has been extensively studied (see, e.g.,
\cite{krokhin04constraint,mei96,nebbur95}). Constraint satisfaction
problems may be viewed as datalog nonemptiness problems for programs
with a single non-recursive rule. Here, we are interested in the
complexity of full datalog over the interval algebra.

\begin{table}[htbp]
  \caption{The 13 basic relations of Allen's interval algebra. The
    obvious inequalities $x^-<x^+$ and $y^-<y^+$ of each case have
    been omitted.} 
  \label{tab:allensalg}
  \centering
  \small
  \begin{tabular}{|ll|ll|c|l|} \hline
    Basic relation && Converse relation && Example & Endpoints \\\hline\hline
    $x$ precedes $y$ & $\mathbf{p}$ &$y$ preceded by $x$ &
    $\mathbf{p}^{-1}$&
    {\tiny$\begin{array}{c}
      xxx\hphantom{xyyyy}\\
      \hphantom{xxxxy}yyy
    \end{array}$}
    & $x^+<y^-$ \\ \hline
    $x$ meets $y$ & $\mathbf{m}$ &$y$ met by $x$ & $\mathbf{m}^{-1}$&
    {\tiny $\begin{array}{c}
      xxxx\hphantom{yyyy}\\
      \hphantom{xxxx}yyyy
    \end{array}$}    
    & $x^+=y^-$ \\ \hline
    $x$ overlaps $y$ & $\mathbf{o}$ &$y$ overlapped by $x$ &
    $\mathbf{o}^{-1}$
    & {\tiny $\begin{array}{c}
      \hphantom{x}xxxx\hphantom{yyyy}\\
      \hphantom{xxx}yyyy\hphantom{yy}
    \end{array}$}    
    &
    $x^-<y^-<x^+<y^+$ \\ \hline
    $x$ during $y$ & $\mathbf{d}$ &$y$ includes $x$ &
    $\mathbf{d}^{-1}$
    &
    {\tiny $\begin{array}{c}
      \hphantom{xyy}xxx\hphantom{yy}\\
      \hphantom{x}yyyyyyy
    \end{array}$}
    &
    $y^-<x^-,\,x^+<y^+$ \\ \hline
    $x$ starts $y$ & $\mathbf{s}$ &$y$ started by $x$ &
    $\mathbf{s}^{-1}$&
    {\tiny $\begin{array}{c}
      xxx\hphantom{yyyyy}\\
      yyyyyyyy
    \end{array}$}    
    & $x^-=y^-,\,x^+<y^+$
    \\\hline
    $x$ finishes $y$ & $\mathbf{f}$ & $y$ finished by $x$ &
    $\mathbf{f}^{-1}$
    & 
    {\tiny $\begin{array}{c}
      \hphantom{yyyyy}xxx\\
      yyyyyyyy
    \end{array}$}    
    & $y^-<x^-,\,x^+=y^+$ \\\hline
    $x$ equals $y$ & $\equiv$ &&&
    {\tiny $\begin{array}{c}
      xxxxx \\
      yyyyy
    \end{array}$}
    & $x^-=y^-,\,x^+=y^+$
        
    \\\hline
  \end{tabular}
\end{table}

Let $\mathcal I$ denote the structure whose universe consists
of all open intervals on the real line, and whose relations are the relations
of the interval algebra. We observe that datalog programs over $\mathcal I$
can easily be translated into programs over the linear order $(\mathbb R,<)$
and vice versa:

\begin{lem}
  The datalog nonemptiness problem over $\mathcal I$ is LOGSPACE-equivalent
  to the datalog nonemptiness problem over $(\mathbb R,<)$.
\end{lem}

\proof
  The reduction from the nonemptiness problem over $\mathcal{I}$ to the one
  over $(\mathbb R,<)$ is straightforward by replacing the interval variables
  by endpoint variables. Since we do not allow any equality relation to be
  used, we simulate equality by identifying variables.
  
  For the other direction, we transform the program $\Pi$ over $(\mathbb
  R,<)$ to $\Pi'$ by replacing all atoms $x<y$ by $\mathbf{p}(x,y)$. Then $\Pi'$
  is satisfiable if and only if $\Pi$ is satisfiable: If $\mathbf{p}(x,y)$
  holds, then $x^-<y^-$ is satisfied. If on the other hand $x<y$ holds, then
  there are elements $x^+$ and $y^+$ such that $\mathbf{p}((x,x^+),(y,y^+))$
  is satisfied, because the order $(\mathbb R,<)$ is dense.

  Both reductions can clearly be carried out in logarithmic space.
\qed

\section{Lower Bounds}
\label{sec:lowbnd}

The hardness results in this section are either
from~\cite{dantsin97complexity}, or they can fairly easily be proved by the
techniques used in~\cite{dantsin97complexity}. It is easy to see that
for every finite structure $\mathcal A$, the datalog
nonemptiness problem over $\mathcal A$ is in EXPTIME. For every structure
$\mathcal A$ whose universe contains at most one element, the nonemptiness
problem is in PTIME.
Conversely, for every structure $\mathcal A$, the datalog nonemptiness
problem over $\mathcal A$ is PTIME-hard, because the
satisfiability problem for propositional Horn clauses is equivalent to the
nonemptiness problem for datalog programs with only $0$-ary relation
symbols. As soon as a structure contains two distinguishable elements, the
nonemptiness problem becomes EXPTIME-hard. This will be made precise in
Lemma~\ref{lem:simulation} below. For the reader's convenience, we sketch the
proof. It requires some preparation.

A \emph{successor structure} is a structure $\mathcal B=(B,S^{\mathcal
  B},N^{\mathcal B})$, where $B$ is either finite or countably infinite, and
for some enumeration $b_0,b_1,\ldots$ of $B$, the binary relation $S^{\mathcal
  B}$ consists of all pairs $(b_{i},b_{i+1})$, and the unary relation
$N^{\mathcal B}$ only contains the element $b_0$.

Assume, that in some structure $\mathcal{A}$ with universe $A$, we can \emph{define} a
successor structure. This means that
there exists a datalog program $\Pi$ with an $m$-ary IDB $U$, a $2m$-ary IDB
$S$, and an $m$-ary IDB $N$ such that the structure $\mathcal B=(B,S^{\mathcal
  B},N^{\mathcal B})$ with $B=U_\infty^{\Pi,\mathcal A}$, $S^{\mathcal
  B}=S_\infty^{\Pi,\mathcal A}$, and $N^{\mathcal B}=N_\infty^{\Pi,\mathcal
  A}$ is a successor structure.
Then a given Turing machine
transition function can be translated to a datalog program defining the
following IDB relations:
\begin{description}
\item[$\mathbf{symbol}_{\sigma}(\bar{x},\bar{y})$] In step $\bar x$ of the
  computation the tape cell $\bar{y}$ contains the symbol $\sigma$.
\item[$\mathbf{cursor}(\bar{x},\bar{y})$] At instant $\bar{x}$ the
  cursor points to cell $\bar{y}$.
\item[$\mathbf{state}_s(\bar{x})$] In step $\bar{x}$ the Turing machine
  is in state $s$. 
\item[$\mathbf{accept}$] The computation has reached an accepting state.
\end{description}
Here $\bar x$ and $\bar y$ range over elements of the defined successor
structure $\mathcal B$ and hence can be viewed as encoding natural numbers,
which are used to address time steps and tape cells.
We may define auxiliary IDB
relations ensuring the consistency of the simulation and encoding the
input on the tape of the machine. Then we have a program, computable
in logarithmic space from the machine encoding, whose IDB
\textit{accept} is derivable (and hence nonempty) if and only if a
machine run accepts in a number of steps bounded by
the size of the successor structure.

\begin{lem}
  \label{lem:simulation}
  Given a structure $\mathcal{A}$ such that two
  relations $U_0, U_1 \subset \mathcal{A}^k$, $k \in \SetN$, can
  be defined by a datalog program on $\mathcal{A}$, such that
  \[
  U_0 \cap U_1 = \emptyset,\; U_0 \ne \emptyset,\; U_1 \ne \emptyset.
  \]
  Then the datalog nonemptiness problem over $\mathcal A$ is EXPTIME-hard.
\end{lem}

\proof
  Without loss of generality we may assume that $\mathcal A$ actually contains
  two $k$-ary relations $U_0,U_1$ which are nonempty and disjoint. Hence we
  can use these relations as EDB predicates in a datalog program.  We prove
  that any deterministic Turing machine computation on input $x$, with $|x|=n$
  and time bound $t(x)=2^m$ ($m=m(n)$ being a function with variable $n$) can
  be simulated by a datalog program with IDBs having at most $2\cdot k \cdot m$
  free variables.

  The elements in $U_0$ are used as 0 and the elements in
  $U_1$ as 1 to build a successor structure of binary vectors of arity $m$,
  leading to a successor substructure with values in $[0..2^m]$. The details
  can be found in \cite{dantsin97complexity} with slight modifications.
  
  The maximal arity of any IDB relations involved is $2\cdot k \cdot
  m$, defining the successor between two $m$-tuples of entries that
  have arity $k$.
  
  By the construction of the Turing machine simulation, any machine
  computation running at most $2^m$ steps can be simulated using datalog
  programs with maximal arity $2\cdot k \cdot m$, which concludes the
  proof.  
\qed

\begin{cor}
  \label{cor:dlexphard}
  For every linear order $\mathcal A=(A,<)$ with at least two elements, the
  datalog nonemptiness problem over $\mathcal A$ is EXPTIME-hard.
\end{cor}

\proof
  Let $U_0$ be the binary relation $x<y$ and $U_1$ the converse $x>y$.
\qed

Note that over infinite structures, the datalog nonemptiness problem can
easily become undecidable. One of the simplest examples of an infinite
structure where this happens is an infinite successor structure:

\begin{prop}
  Let $\mathcal B$ be an infinite successor structure. Then the datalog
  nonemptiness problem over $\mathcal B$ is undecidable. \qed
\end{prop}
The proof is another straightforward Turing machine simulation.

\section{Distance types}
\label{sec:types}

Types are a model theoretic tool that we shall use  for dealing with datalog
programs on infinite orders. We define an appropriate notion of type and prove
a lemma that links them with the evaluation of datalog
programs. 

\begin{defi}\hfill
  \begin{enumerate}[(1)]
  \item A \emph{distance atom} is an expression of the form $x\le_d
    y$, $-\infty\le_d x$, or $x\le_d\infty$, where $x,y$ are variables and
    $d$ is a nonnegative integer. We may write $<_d$ instead of
    $\le_d$ for $d>0$. A \emph{distance type} is a finite set of
    distance atoms.
    
    We write $\delta(x_1,\ldots,x_k)$ to indicate that the variables of the
    distance type $\delta$ are among $x_1,\ldots,x_k$. The set of all distance
    types with variables among $x_1,\ldots,x_k$ is denoted by
    $\Delta(x_1,\ldots,x_k)$.
  \item Let $\mathcal A=(A,<)$ be a linear order, $\bar a=(a_1,\ldots,a_k)\in
    A^k$, and let $\delta=\delta(x_1,\ldots,x_k)$ be a distance type. Then
    $(\mathcal A,\bar a)$ \emph{satisfies} $\delta$  (we write: $\mathcal
    A\models\delta(\bar a)$),\footnote{Another common terminology is to say
    that a type is \emph{``realised''} instead of ``satisfied''.} if 
    \begin{enumerate}[$-$]
    \item for all atoms $x_i\le_d x_j\in\delta$, there
      are $b_0,\ldots,b_d\in A$ such that $a_i\le b_0<b_1<\ldots<b_d\le a_j$ (that
      is, $x_i\le_d x_j$ is interpreted as $x_i\le x_j$ and $d(x_i,x_j)\ge d$);
    \item for all atoms $-\infty\le_d x_j\in\delta$, there
      are $b_0,\ldots,b_d\in A$ such that $b_0<b_1<\ldots<b_d\le a_j$;
    \item for all atoms $x_i\le_d \infty\in\delta$, there
      are $b_0,\ldots,b_d\in A$ such that $a_i\le b_0<b_1<\ldots<b_d$.
    \end{enumerate}
    A distance type $\delta$ is \emph{satisfiable} if there is a linear order
    $\mathcal A$ and a tuple $\bar a$ such that $(\mathcal A,\bar a)$
    satisfies $\delta$.
  \item The \emph{rank} of a distance atom $t\le_d u$ is $d$, and the rank of
    a distance type $\delta$ is the maximum of the ranks of all atoms it
    contains. The set of all distance types in $\Delta(x_1,\ldots,x_k)$ of
    rank at most $d$ is denoted by $\Delta_d(x_1,\ldots,x_k)$.
  
  \item Let $\mathcal A=(A,<)$ be a linear order, $\bar a=(a_1,\ldots,a_k)\in
    A^k$, and $d\ge0$. The \emph{distance-$d$ type of $\bar a$ in $\mathcal
      A$}, denoted by $\tp_d(\mathcal A,\bar a)$, is the distance type that
    contains:
    \begin{enumerate}[$\bullet$]
    \item for $1\le i,j\le k$ with $a_i\le a_j$ the distance atom $x_i\le_c
      x_j$, where $c=\min\{d,d(a_i,a_j)\}$;
    \item for $1\le j\le k$ the distance atom $-\infty\le_c x_j$, where $c\le d$
      is maximum such that there exists $b_0,\ldots,b_c\in A$ with
      $b_0<\ldots<b_c\le a_j$;
    \item for $1\le i\le k$ the distance atom $x_i\le_c\infty$, where $c\le d$
      is maximum such that there exists $b_0,\ldots,b_c\in A$ with
      $a_i\le b_0<\ldots<b_c$.
    \end{enumerate}

  \item A distance type $\delta$ is \emph{complete} if there exists a
    linear order $\mathcal A$, a tuple $\bar{a}$ with
    $\mathcal{A}\vDash \delta(\bar{a})$, and $d\ge 0$ such that for
    each pair $(a_i,a_j)$ of entries of $\bar{a}$ satisfying $a_i<a_j$
    there is precisely one distance atom $a_i\le_c a_j$ in $\delta$ with $0<c\le
    d$, and for each pair $(a_i,a_j)$ with $a_i=a_j$ there are
    distance atoms $a_i\le_0 a_j$ and $a_j\le_0 a_i$ in $\delta$.
    
    The set of all complete distance types with variables among
    $x_1,\ldots,x_k$ is denoted by $\Gamma(x_1,\ldots,x_k)$, and the
    set of all types in $\Gamma(x_1,\ldots,x_k)$ of rank at most $d$
    is denoted by $\Gamma_d(x_1,\ldots,x_k)$.
  \end{enumerate}
\end{defi}

\begin{exa}\label{ex:disttype}
  An example for a distance type from $\Delta(x,y,z)$ is:
  \[
  \delta = x<_3 y,\; y<_2 z
  \]
  This type $\delta$ is satisfied for some elements $a_1,a_2,a_3\in
  A$, which we assign to the variables $x= a_1$, $y= a_2$ and
  $z= a_3$, if there exist $b_1,b_2,b_3\in A$ with
  \begin{align*}
    a_1< b_1 <b_2 < a_2 &&& \text{to satisfy } x<_3 y \\
    a_2< b_3 <a_3 &&& \text{to satisfy } y<_2 z\enspace.
  \end{align*}
  The occurring ranks of the atoms in delta show $\delta \in
  \Delta_3(x,y,z)$. $\delta$ is not complete, since there is no
  distance atom containing $x$ and $z$ and no distance atom containing
  $-\infty$ or $\infty$.
\end{exa}

Let us point out some subtleties of these definitions that may be
confusing. A distance type need not be satisfiable, but a complete distance
type must be satisfiable.\footnote{In model theory, it is common to define
  types as being satisfiable sets of formulas.} Even though the ``constants'' $-\infty$ and $\infty$ appear in distance
atoms, they are not part of the datalog language, and we do not require linear
orders to have a minimum or maximum. The semantics of the atoms $-\infty\le_d x$,
or $x\le_d\infty$ is well-defined in all linear orders.

Note that $x\le_1 y$ is equivalent to $x<y$ and that $x\le_0 y\wedge y\le_0 x$
is equivalent to $x=y$. A distance type of rank $1$ only contains information
about the relative order of the variables and about equalities between the
variables, and not about their distances. Hence we call distance types of rank
$1$ \emph{order types}.

It it is easy to see that it can be decided in polynomial time in the number of
variables whether a distance type is satisfiable and whether it is complete.

\begin{defi}
  Let $\gamma,\delta$ be distance types. Then $\gamma$ \emph{implies}
  $\delta$ if for all distance atoms $x\le_d x'$ in $\delta$ there is
  a distance atom $x\le_{d'}x'$ with $d\le d'$ in $\gamma$.  
\end{defi}

\begin{lem}\label{lem:type-implication}\hfill
  \begin{enumerate}[\em(1)]
  \item Let $\gamma(\bar x),\delta(\bar y)$ be distance types such
    that $\gamma$ implies  $\delta$. Then all variables in $\bar y$
    also appear in $\bar x$, and for every linear order $\mathcal A$
    and every tuple $\bar a$ such that $\mathcal A\models\gamma(\bar
    a)$, for the projection $\bar b$ of $\bar a$ to the variables in
    $\bar y$ we have $\mathcal A\models\delta(\bar b)$.
  \item Let $\delta\in\Delta_d(\bar x)$. Then for all linear orders
    $\mathcal A$ and all tuples $\bar a\in A^k$,
    \begin{equation}\label{eq:disj}
      \mathcal A\models\delta(\bar a)\iff\text{there exists a
        $\gamma\in\Gamma_d(\bar x)$ such that $\gamma$ implies $\delta$ and $\mathcal A\models\gamma(\bar a)$.}
      \end{equation}
    \end{enumerate}
\end{lem}
We omit the straightforward proof. Note that
statement (2) of the lemma implies that every type can be written as a
disjunction of complete types. 

Recall that for each IDB $P$ of a datalog program $\Pi$ over some fixed
structure $\mathcal A$, by $P_i^{\Pi}$ we denote the interpretation of $P$
after the $i$th stage of the computation of $\Pi$. In the following
lemma we will show how to describe the stages by finite sets of
distance types, but first we will have a look at a simple example.

\begin{exa}\label{ex:disttypeanddl}
  Let $\Pi$ be the following two-rule program defining IDBs $P$ and $Q$:
  \begin{align*}
    &P(x,y) \rd x<z_1,\, z_1<z_2,\, z_2<y. \\
    &Q(x,y,z) \rd P(x,y),\, y<w,\, w<z.
  \end{align*}
  Applying the first rule to the empty IDB relations at the beginning,
  the resulting relation $P^{\Pi}_1$ contains all tuples satisfying
  the distance type $x<_3 y$, since there have to be three distance
  atoms satisfied between the elements assigned to $x$ and $y$.
  
  Applying the second rule to this stage 1, this distance type is
  copied to the type describing the tuples in $Q^{\Pi}_2$ and on $y$
  and $z$ the type $y<_2 z$ is imposed, leading to the following type
  describing $Q^{\Pi}_2$:
  \[
  \delta = x<_3 y,\; y<_2 z
  \]
\end{exa}

For programs using recursion and more rules leading to some form of
disjunction, a single distance type is not enough to describe a
relation, but sets of types are needed.

\begin{lem}
  \label{lem:discrtypes}
  Let $\mathcal{A}=(A,<)$ be an infinite linear order and $\Pi$ a datalog
  program over $\mathcal A$.
 
  Then for each $k$-ary IDB $P$ of $\Pi$ and each $i\ge 0$ there is a
  finite set $\Theta(P,i)\subseteq
  \Gamma(x_1,\ldots,x_k)$ such that for all $\bar a\in A^k$ it holds
  that
  \[
  \bar a\in P_i^\Pi\iff\text{there is a $\theta\in \Theta(P,i)$ such that
  }\mathcal A\models\theta(\bar a).
  \]
  Furthermore, the rank of all types in $\Theta(P,i)$ is bounded by $(m_R)^i$,
  where $m_R$ denotes the maximal number of variables in a rule of $\Pi$ as
  usual.
\end{lem}

\proof
  For $i\ge 0$, let $d_i:=(m_R)^i$.

  We prove the claim by induction on $i$. The induction base for
  $i=0$ is obvious: We let $\Theta(P,0)=\emptyset$ for all IDBs $P$.
  
  For the induction step ($i\to i+1$), we consider the application of a rule
  \begin{align*}
    \rho:\;P(\bar x) &\rd 
    P^1(\bar y_1),\ldots,P^\ell(\bar y_\ell),\epsilon(\bar y),
  \end{align*}
  where $P^1,\ldots,P^\ell$ are IDBs and $\epsilon(\bar y)$ is a list
  of EDB atoms with variables in $\bar y$. We view $\epsilon$ as a
  distance type of rank 1; this will enable us to unify some of the
  arguments below. Let $Z=\{z_1,\ldots,z_m\}$ be the set of all
  variables occurring in $\rho$, and let $\bar
  z=(z_1,\ldots,z_m)$. Note that $m\le m_R$ and hence $m\cdot d_i\le
  d_{i+1}$.
  
  For $1\le j\le \ell$, let $\theta_j\in\Theta(P^j,i)$. Assume that
  there is a $\gamma\in\Gamma_{d_i}(z_1,\ldots,z_m)$ which implies
  $\theta_1,\ldots,\theta_\ell$ and $\epsilon$. 
  Suppose $w_1,\ldots,w_m$ is an enumeration of $Z$ in the order
  imposed by $\gamma$, and let $e_0,\ldots,e_{m}\ge 0$ such that
  $\gamma$ contains the distance atoms
  \[
  -\infty\le_{e_0}w_1,\quad w_{p}\le_{e_{p}}w_{p+1}\text{ for
  }p=1,\ldots,m-1,
  \quad
  w_m\le_{e_m}\infty.
  \]
  We define a new type $\gamma|_{\bar x}$ in the variables $\bar x$ as follows: 
  \begin{enumerate}[$\bullet$]
  \item Let $x$ be a variable in $\bar x$, say, $x=w_p$. Then 
    $\gamma|_{\bar x}$ contains the distance atom $-\infty\le_d x$ for
    $d=\sum_{q=0}^{i-1}e_q$.
  \item Let $x,x'$ be variables in $\bar x$, say, $x=w_i$ and
    $x'=w_j$. Suppose that $i\le j$. Then 
    $\gamma|_{\bar x}$ contains the distance atom $x\le_d x'$ for
    $d=\sum_{r=p}^{q-1}e_r$.
  \item Let $x$ be a variable in $\bar x$, say, $x=w_p$. Then 
    $\gamma|_{\bar x}$ contains the distance atom $x\le_d\infty$ for
    $d=\sum_{q=p}^{m}e_q$.
  \end{enumerate}
  It is easy to see that $\gamma|_{\bar x}$ is a complete distance
  type in the variables $\bar x$. The rank of $\gamma|_{\bar x}$ is at
  most $m$ times the rank of $\gamma$ and hence bounded by $m\cdot
  d_i\le d_{i+1}$. Therefore, $\gamma|_{\bar
    x}\in\Gamma_{d_{i+1}}(\bar x)$. Furthermore, it is easy to verify
  that every tuple $\bar a$ that satisfies $\gamma|_{\bar x}$ has an
  extension to an $m$-tuple $\bar c$ that satisfies $\gamma$. 

  We let $\Theta(P,i+1)$ be the union of $\Theta(P,i)$ with all types
  $\gamma|_{\bar x}$, where $\gamma\in\Gamma_{d_i}(\bar z)$ such that
  $\gamma$ implies $\epsilon$ and there exist
  $\theta_j\in\Theta(P^j,i)$, for $1\le j\le \ell$, implied by $\gamma$. We claim that for all tuples $\bar a$ it holds that
  $\bar a\in P^\Pi_{i+1}$ if and only if there is a
  $\theta\in\Theta(P,i+1)$ such that $\mathcal A\models\theta(\bar
  a)$.

  To prove the forward direction of this claim, let $\bar a\in
  P^\Pi_{i+1}$. If $\bar a\in P_i^\Pi$, then by the induction
  hypothesis there is a $\theta\in\Theta(P,i)\subseteq\Theta(P,i+1)$
  such that $\mathcal A\models\theta(\bar a)$. Suppose that $\bar a\in
  P^\Pi_{i+1}\setminus P_i^\Pi$. Then there is a tuple $\bar c$
  interpreting the variables $\bar z$, with projections $\bar a$ on
  the coordinates of 
  the variables in $\bar x$, $\bar b_j$ on
  the coordinates of the variables in $\bar
  y_j$, and $\bar b$ on
  the coordinates of the variables in $\bar y$, such that $\bar
  b_j\in (P^j)^\Pi_i$ for $1\le j\le \ell$ and $\mathcal A\models\epsilon(\bar b)$. By the induction hypothesis, for $1\le j\le\ell$ there
  is a type $\theta_j\in\Theta(P_j,i)$ such that $\mathcal
  A\models\theta_j(\bar b_j)$. Let $\gamma$ be the complete
  distance-$d_i$ type of $\bar c$. Then $\gamma|_{\bar
    x}\in\Theta(P,i+1)$, and $\mathcal A\models\gamma|_{\bar x}(\bar a)$.

  For the backward direction, suppose that a tuple $\bar a$ satisfies
  a type $\gamma'(\bar x)\in\Theta(P,i+1)$. If $\gamma'(\bar
  x)\in\Theta(P,i)$, then $\bar a\in P^\Pi_i\subseteq P^\Pi_{i+1}$ by
  the induction hypothesis. Otherwise, there is a complete type
  $\gamma\in\Gamma_{d_i}(\bar z)$ and types $\theta_j\in\Theta(P^j,i)$
  for $1\le j\le \ell$ such that $\gamma'=\gamma|_{\bar x}$ and
  $\gamma$ implies
  $\theta_1,\ldots,\theta_\ell,\epsilon$. Let $\bar c$ be an $m$-tuple
  satisfying $\gamma$ such that the projection of $\bar c$ on the
  coordinates of the variables in $\bar x$ is $\bar a$. For $1\le
  j\le\ell$, let $\bar b_j$ be the projection of $\bar c$ on the
  the
  coordinates of the variables in $\bar y_j$. Then $\mathcal A\models\theta_j(\bar
  b_j)$. By the induction hypothesis, $\bar b_j\in
  (P^j)^\Pi_i$. Let $\bar b$ be the projection of $\bar c$ on the
  coordinates of the variables in $\bar y$. Then $\mathcal
  A\models\epsilon(\bar b)$. Putting everything together, we obtain
  $\bar a\in P^\Pi_{i+1}$.
\qed

\begin{rem}\label{rem:rank}
  Note that actually we have proved a slightly stronger bound on the
  ranks of the types in $\Theta(P,i)$: Letting $d_i$ be the maximum
  rank of all types in $\Theta(P,i)$, we have 
  \begin{equation}
    \label{eq:rank}
    d_{i+1}\le
  m_R\cdot d_i
  \end{equation}
  for all $i\ge0$. Furthermore, whereas the numbers $d_i$ may depend
  on the order in which we apply the rules of the program, the bound
  \eqref{eq:rank} holds for all orders.
\end{rem}

The following example shows that the ranks
of the types can increase during a computation in a way that can get quite
complicated:

\begin{exa}
  \label{ex:discrrun}
  Consider the following program consisting of rules $\rho_1$,
  $\rho_2$ and $\rho_3$, with $\bar{x}=(x_1,\dotsc,x_5)$. We use the
  abbreviation $x_i <_2 x_j$ for $x_i < y, y < x_j$ omitting some body
  variable $y$ in the definition of $\Pi$, which does not appear
  elsewhere in the rules.
  \begin{align*}
    \rho_1: P\bar{x}\rd& x_1<_2 x_2,\, x_2<_2 x_3,\, x_4<_2 x_5. \\
    \rho_2: P\bar{x}\rd& x_1 < x_2,\, x_4<z_2,\, z_3<y_4,\, y_5<x_5,\,\\
    & P(x_2,x_3,z_1,z_2,z_3),\,P(y_1,y_2,y_3,y_4,y_5).\\
    \rho_3: P\bar{x}\rd&
    P(x_1,x_2,x_3,z_1,z_2),P(y_1,x_4,x_5,y_2,y_3). 
  \end{align*}
  The rule $\rho_1$ is an initialization rule which initializes all given
  types to $<_2$. 

  The rule $\rho_2$ introduces $x_1 <_1 x_2$, reuses
  existing types by copying and sums up some existing atoms from possibly
  different existing types. 
  
  This rule uses two recursive occurrences of the IDB $P$, which in
  our description of the application of this rule by types leads to
  the use of two (possibly different) types from the type set
  describing earlier stages of $P^{\Pi}_{\infty}$. We denote the ranks of
  the distance atoms from these two types occuring in our computation,
  using $\bar{a}=(a_1,\dotsc,a_5)$ for tuples from such stages, by:
  \begin{center}
    \begin{tabular}{|l|l|l|} \hline
      \multicolumn{3}{|l|}{Ranks in distance types of the earlier
      stages of $P^{\Pi}_{\infty}$} 
      \\\hline\hline
      first occurrence of $P$ & $a_1<_{c_{1}}a_2$ & $a_4<_{c_{2}}a_5$
      \\\hline 
      second occurrence of $P$ & $a_1<_{c'_{1}}a_2$ &
      $a_4<_{c'_{2}}a_5$ \\\hline 
    \end{tabular}
  \end{center}
  
  The rule application of $\rho_2$ using these distance atoms will
  then impose the following type on the body variables, where the
  distance atoms are given in the order of the variable appearance in
  the rule, ommiting the atoms containing the variables $z_1$, $y_1$,
  $y_2$ and $y_3$, not part of the result:
  \[
  x_1<_1 x_2,\, x_4<z_2,\, z_3 <_1 y_4,\, y_5<_1 x_5,\, x_2<_{c_1}x_3,\,
  z_2<_{c_2}z_3,\, y_4<_{c'_2}y_5
  \]

  To combine these types by eliminating non-head variables, we
  rearrange these atoms:
  \[
  x_1<_1 x_2,\, x_2<_{c_1}x_3,\,x_4<_1 z_2,\, z_2<_{c_2}z_3,\,
  z_3 <_1 y_4,\, y_4<_{c'_2}y_5,\, y_5<_1 x_5
  \]
  
  After the elimination of non-head variables, the following type is
  added to the type set of $P$:

  \[
  x_1<_1 x_2,\, x_2<_{c_1}x_3,\,x_4<_{c_2+c'_2+3} x_5
  \]

  Rule $\rho_3$ copies some distance atoms for $x_1,x_2,x_3$ and
  transfers some $x_2<_c x_3$ to $x_4<_c x_5$ in the result.  We
  conclude the example with the shortest program run leading to a
  fixed point, described by the ranks of the types $x_1<_{d_1} x_2$,
  $x_2<_{d_2} x_3$ and $x_4<_{d_3} x_5$. We assume, that always the
  smallest ranks are chosen.  Longer runs could lead to even bigger
  intermediate results, but will have the same final result.
  \begin{center}
    \small
    \begin{tabular}{|c|l|c|c|c|l|} \hline
      step & rule & $d_1$ & $d_2$ & $d_3$ & remarks \\\hline\hline
      1 & $\rho_1$ & 2 & 2 & 2 & \\\hline
      2 & $\rho_2$ & 1 & 2 & 7 & \\\hline
      3 & $\rho_2$ & 1 & 1 & 12 & using tuples in line 2 and 1 \\\hline
      4 & $\rho_3$ & 1 & 1 & 1 & using tuple in line 3 twice \\\hline
    \end{tabular}
  \end{center}
\end{exa}

\section{Upper Bounds for datalog programs on orders}
\label{sec:uppbnd}

Now that we have a formal description of the IDB relations in this
case, we will use the concept of discrete order types to show an upper
bound for the datalog nonemptiness problem. But before, we transform the
program into some normal form which integrates the possible order
types into the program by creating disjoint copies of each IDB,
each having a different order type and hence leading to disjoint
relations. 

\begin{defi}
  A datalog program over linear orders $\mathcal A$ is \emph{type-disjoint},
  if for every $k$-ary IDB $P$ there is a complete order type
  $\gamma_P\in\Gamma_1(x_1,\ldots,x_k)$ such that for all linear orders
  $\mathcal A=(A,<)$ and all tuples $\bar a\in P_\infty^\Pi$ it holds that
  $\tp_1(\mathcal A,\bar a)=\gamma_P$.
  
  The \emph{order type} of an IDB $P$ in a type-disjoint program $\Pi$ is the
  order type $\gamma_P$.
\end{defi}

\begin{lem}
  \label{lem:disjoint}
  For every datalog program $\Pi$ over linear orders there is a type-disjoint
  datalog program $\Pi'$ over linear orders with the following properties:
  \begin{enumerate}[\em(1)]
  \item For every IDB $P$ of $\Pi$ there are IDBs $P_1,\ldots,P_{n_P}$ of
  $\Pi'$ of pairwise distinct order types, such that for every linear order
  $\mathcal A$,
  \[
  P_\infty^{\Pi}=\bigcup_{j=1}^{n_P}(P_j)_\infty^{\Pi'}.
  \]
\item $n'_I\le n_I\cdot 3^{m_L^2}$, $n'_R\le 3^{m_L^2\cdot (m'_I+1)}\cdot
  n_R$, $m'_R=m_R$ and $m'_L=m_L$, $m'_I=m_I$, where
  $n'_I$, $m'_R$, $m'_L$, $m'_I$ are the parameters of $\Pi'$.
  \end{enumerate}
  Furthermore, the program $\Pi'$ can be computed from $\Pi$ in exponential
  time.
\end{lem}
\proof From each IDB $P$ of arity $r$, we create $n_P=3^{r^2}$ distinct
copies $P_0,\dotsc,P_{n_P-1}$, each having a different order type. For
$i\in\{0,\dotsc,n_P-1\}$, let $(i_0,\dotsc,i_{r^2-1})$ be the ternary
representation of the number $i$, $i=\sum_{j=0}^{r^2-1} i_j \cdot 3^j$
with $0\le i_j < 3$ for all $j=1,\dotsc,n_P-1$. Then we link to each
new IDB $P_i$ a distance-1-type $\gamma_{P_i}$, which consists of the
following distance atoms:
\begin{align*}
  x_{j_1} = x_{j_2},\quad & \text{ if }\; i_{j_1+j_2\cdot r} = 0 \\
  x_{j_1} <_1 x_{j_2},\quad &\text{ if }\; i_{j_1+j_2\cdot r} = 1 \\
  x_{j_2} <_1 x_{j_1},\quad &\text{ if }\; i_{j_1+j_2\cdot r} = 2 
\end{align*}

So each combination of distance atoms for all pairs of variables will
be present in some $\gamma_{P_i}$.  After computing these distance
types, we transform the program in two stages. First, we change the
head IDBs to the new IDB set consisting of the distinct copies created
as above for each IDB of $\Pi$: Each rule $\rho$ with head atom
$P\bar{x}$, is replaced by copies $\rho'_0,\dotsc,\rho'_{n_P-1}$ with
head $P_j\bar{x}$ with $j\in\{0,\dotsc,n_P-1\}$, and the body copied
from $\rho$ and extended by EDBs $x_{j_1}<x_{j_2}$ for each
$(x_{j_1}<_1 x_{j_2})\in \gamma_{P_j}$.  In case of $(x_{j_1}=x_{j_2})
\in \gamma_{P_j}$, we replace all occurrences of $x_{j_2}$ by
$x_{j_1}$ afterwards. This simulates the equality relation, which is
not available as EDB or IDB relation.

Each of the rules $\rho'_0,\dotsc,\rho'_{n_P-1}$ is then itself replaced
by copies which instead of the body IDBs from $\Pi$, use the IDBs of
$\Pi'$:

In each rule $\rho_r$, say, $P_l\bar{x}\rd
Q_1\bar{y}_1,\dotsc,Q_{m}\bar{y}_{m},\epsilon(\bar{x},\bar{y}_1,\dotsc,\bar{y}_m)$,
with $\epsilon$ being a sequence of EDBs, each IDB $Q_j$ of $\Pi$ has been
converted to a set of IDBs $\{Q_{j\ell}\}$. From these sets we
generate all possible combinations $(Q_{1\ell_1}, \dotsc,
Q_{m\ell_m})$ and create from each combination a rule of $\Pi'$:
\begin{align*}
  P_l\bar{x}\rd
  Q_{1\ell_1}\bar{y}_1,\dotsc,Q_{m\ell_m}\bar{y}_m,\;
  \epsilon(\bar{x},\bar{y}_1,\dotsc,\bar{y}_m)\enspace,
\end{align*}
where the sequence $E$ of EDB atoms is left untouched.

After that, we directly eliminate a rule with an inconsistent order
type. This can be done by viewing the rule as graph with the variables
being the nodes and the order atoms being directed edges. A check for
a directed cycle, which can be carried out in time polynomial in the
rule length, shows if the order type is inconsistent.

Each tuple added to a stage in the evaluation of $\Pi$ introduced by
some rule $\rho$ of $\Pi$ has a complete distance 1 type, so there
will be one of the copies of $\rho$, which can be applied to add this
tuple. Conversely, the newly created rules of $\Pi'$ may only add
tuples, for which a rule in $\Pi$ exists adding this tuple.

Each IDB of arity $r$ is converted to not more than $3^{r^2}$ copies and
hence $n'_I\le n_I\cdot 3^{m_L^2}$. For each rule, we need all
combinations of copies of the newly created IDBs, adding up to at most
$n'_R\le \left( 3^{m_L^2} \right)^{(m_I+1)}\cdot n_R = 3^{m_L^2\cdot
  (m_I+1)}\cdot n_R$. \qed

For type-disjoint datalog programs, the nonemptiness problem can be solved in
a simple fashion, essentially disregarding any recursion in rules. In
the following lemma, we construct an execution sequence $s$ that will suffice
to decide the datalog nonempti\-ness problem.

\begin{lem}
  \label{lem:discinit}
  Let $\Pi$ be a type-disjoint datalog program over an infinite linear order
  $\mathcal{A}=(A,<)$. Then there exist an $i_s\le n_I$ and a
  sequence $s=(\rho_0,\rho_1,\dotsc,\rho_{i_s-1})$ of rules,
  such that after applying $\rho_i$ to stage $i$ for $i=0,\dotsc,i_s-1$,
  the emptiness is determined, i.e. for all IDBs $P$ it holds that
  \begin{align}
    \label{eq:discinit}
    P_{i_s}^{\Pi}=\emptyset\quad\Rightarrow\quad
    P_{\infty}^{\Pi}=\emptyset \enspace.
  \end{align}
  Such a sequence $s$ can be computed in time $n_R\cdot n_I$.
\end{lem}

\proof
  We create the sequence $s$ by cycling through the rules $n_I$ times,
  adding those rules to $s$, which change an empty IDB to nonempty,
  formally: $s=(\rho_0,\rho_1,\dotsc,\rho_{i_s-1})$ such that there
  exist IDBs $P_1,\dotsc,P_{i_s}$ with $(P_i)^{\Pi}_i=\emptyset$, and
  after applying $\rho_i$ to $(P_i)^{\Pi}_i$,
  $(P_i)_{i+1}^{\Pi}\ne\emptyset$ for $i=0,\dotsc,i_s-1$.
  
  We continue this process until no more rules can be applied to make
  an empty IDB nonempty, but this can happen at most $n_I$ times,
  immediately leading to the time bound for the computation. Note that
  nonempty IDBs are never modified by $s$. 
  
  The crucial observation is that, in a type-disjoint program, it only
  depends on the nonemptiness of the IDBs in the body if a rule adds
  new tuples to the head IDB, and not on the actual content of the
  body IDBs. This follows from the fact that at each stage the content
  of the IDBs is a union of disjoint complete types by
  Lemma~\ref{lem:discrtypes}, and that the distance types are monotone
  by Corollary~\ref{cor:types1}. Thus in an infinite order, we can always
  add all tuples of sufficiently large finite distances.
  
  We now show property (\ref{eq:discinit}) by contradiction:
 
  Let $U=\condset{R}{R^{\Pi}_{i_s}=\emptyset\; \wedge\;
    R^{\Pi}_{\infty}\ne\emptyset}$ be the set of IDBs changing to
  nonempty after $s$ and assume $U\ne\emptyset$. Then for each $R\in U$
  there exist an $i_R\in\SetN$ and a rule $\rho_R$ with:
  \begin{align*}
  R^{\Pi}_{i_R}=\emptyset, \text{ and applying } \rho_R 
  \text{ to } R^{\Pi}_{i_R}:\; R^{\Pi}_{i_R+1}\ne\emptyset\enspace.  
  \end{align*}
  Let $P \in U$ be the IDB with $i_P=\min\condset{i_R}{R\in U}$. By
  the definition of $U$ and by the choice of $i$, all $Q \in
  U\setminus \{P\}$ have to satisfy $Q^{\Pi}_{i_R}=\emptyset$. Since a
  rule can be applied if and only if all body IDBs are nonempty, the
  rule $\rho_R$ cannot depend on them and can be applied in stage
  $i_s$ leading to a sequence of rule applications making more IDBs
  nonempty, a contradiction to the construction of $s$.
  \qed

\begin{thm}
  \label{thm:discrtime}
  The datalog nonemptiness problem over linear orders $\mathcal{A}=(A,<)$
  is EXP\-TIME-complete
\end{thm}

\proof
  The proof is a combination of several earlier results. A datalog program
  $\Pi$ can by Lemma \ref{lem:disjoint} be converted to a
  type-disjoint program $\Pi'$. For this kind of program Lemma
  \ref{lem:discinit} gives us a method to check which IDB relations of
  $\Pi'$ will be empty after an evaluation of $\Pi'$. Since $\Pi'$ is
  type-disjoint, each IDB relation of the original program $\Pi$ will
  occur here as a collection of IDBs of $\Pi'$, which can easily be
  determined. Thus, the question ``$P^{\Pi}_{\infty}=\emptyset$?'' can
  be answered by checking the type sets of all corresponding IDBs of
  $\Pi'$. Beside the time for this check and the time for the
  conversion of the programs, the time for determining the empty IDB
  relations of $\Pi'$ is part of the running time. Using Lemma
  \ref{lem:disjoint} and \ref{lem:discinit} the time of this step can
  be bounded from above by $\bigoh{n_R\cdot 9^{m_L^2\cdot (m_I+1)}}$,
  altogether clearly in EXPTIME and with the earlier shown
  EXPTIME-hardness the claim follows.  \qed

\section{Boundedness}
A datalog program $\Pi$ is \emph{bounded on a structure $\mathcal A$}
if there is computation of $\Pi$ on $\mathcal A$, that reaches a fixed
point after finitely many stages. Of course, this concept of
boundedness is nontrivial only on infinite structures. The main result
of this section is that datalog programs are bounded on linear orders.
Actually, we prove a stronger result giving a uniform bound on the
number of evaluation steps that computable from the size of the
program and does not depend on the structure. This stronger result is
even meaningful for finite linear orders.

\begin{defi}
  Let $\Pi$ be a datalog program over a structure $\mathcal A$.
  \begin{enumerate}[(1)]
  \item A \emph{computation sequence} for $\Pi$ is a sequence $s$ of
    rules of $\Pi$ to compute all IDB relations, i.e. a sequence of
    rules satisfying the following conditions:
    \begin{enumerate}[$\bullet$]
    \item If $s$ is finite, then after applying $s$, no further rule
      application adds a tuple to the IDB relations. 
    \item If $s$ is infinite, then each rule of $\Pi$ will occur
      infinitely often.
    \end{enumerate}
  \item The \emph{closure ordinal} of $\Pi$ on $\mathcal A$, denoted
    by $\cl(\Pi,\mathcal A)$, is the length of the shortest
    computation sequence for $\Pi$ on $\mathcal{A}$ ($\cl(\Pi,\mathcal
    A)=\infty$, if all computation sequences are of infinite length).
  \item $\Pi$ is \emph{bounded on $\mathcal A$} if $\cl(\Pi,\mathcal
    A)<\infty$.
  \end{enumerate}
  Now let $C$ be a class of structures such that $\Pi$ is a program over $C$.
  \begin{enumerate}[(1)]
    \setcounter{enumi}{3}
  \item The \emph{uniform closure ordinal} of $\Pi$ on $C$, denoted by
    $\ucl(\Pi,C)$, is the maximum of the closure ordinals $\cl(\Pi,\mathcal
    A)$ for $\mathcal A\in C$, if this maximum exists,
    and $\infty$ otherwise.
  \item $\Pi$ is \emph{uniformly bounded on $C$} if $\ucl(\Pi,C)<\infty$.
  \end{enumerate}
\end{defi}

Note that if $\Pi$ is uniformly bounded on $C$, then it is bounded on all
$\mathcal A\in C$, but that the converse does not necessarily hold.

\begin{thm}\label{theo:ucl}
  Let $\Pi$ be a datalog program over the class LO of linear orders. Then $\Pi$
  is uniformly bounded on LO. More precisely, there is a computable
  function $b:\SetN\mapsto\SetN$ such
  that for $n=|\Pi|$ it holds that 
  \[
  \ucl(\Pi,\text{LO})\le b(n).
  \]
\end{thm}

Our proof of Theorem~\ref{theo:ucl} is based on a simplification of
the distance type concept which we will discuss before the
presentation of the main proof. The proof presented here is an
extension of the proof of Theorem~\ref{thm:discrtime}, first
transforming the program $\Pi$ in question to a type-disjoint version
$\Pi'$ by Lemma~\ref{lem:disjoint} and then creating the
initialization sequence $s$ as in Lemma~\ref{lem:discinit}. After this
process, we may eliminate all then empty IDB relations. 
Each remaining IDB $P$ may only contain tuples of one complete order
type $\vartheta_P$.

Let $\gamma\in\Gamma(x_1,\ldots,x_n)$ be a complete distance
type. Observe that $\gamma$ is completely determined by its underlying
order type and the distances $d$ imposed by the distance atoms
$-\infty\le_d x$, $x\le_d x'$, $x\le_d \infty$. We can describe the
distances by a tuple $\bar d^\gamma=(d_1^\gamma,\ldots,d_k^\gamma)$ of
length $k=2n+\binom{n}{2}$ with nonnegative integer entries. We call
$\bar d^\gamma$ the \emph{rank vector} of $\gamma$. We define a
partial order $\preceq$ on the complete distance types in
$\Gamma(x_1,\ldots,x_n)$ by letting $\gamma\preceq\gamma'$ if $\gamma$
and $\gamma'$ imply the same order type and $d_i^\gamma\le
d_i^{\gamma'}$ for $1\le i\le k$. Observe that $\gamma\preceq\gamma'$
if and only if $\gamma'$ implies $\gamma$. The following corollary is
hence an immediate consequence of Lemma~\ref{lem:type-implication}(1):

\begin{cor}\label{cor:types1}
  Let $\mathcal A=(A,<)$ be a linear order, $\bar a\in A^k$, and
  $\gamma,\gamma'\in\Gamma(x_1,\ldots,x_k)$ such that $\gamma\preceq\gamma'$. Then
  \[
  \mathcal A\models\gamma'(\bar a)\implies\mathcal A\models\gamma(\bar a).
  \]
\end{cor}

The crucial observation that we will exploit in the following is that the
computation of a type-disjoint datalog program can be described
entirely in terms of sequences of rank vectors for the IDBs. This
follows from Lemma~\ref{lem:discrtypes} stating that the computation can be
described in terms of complete types and the observation that for 
type-disjoint programs it suffices to consider the rank vectors, because
the order types of the IDBs are fixed.

After applying the sequence $s$ and after eliminating empty IDBs, for
each IDB $P$, the set $\Theta(P,i_s)$ is described by exactly one such
vector, since $|\Theta(P,i_s)|=1$ after the initialization sequence
which adds at most one type to the type set of each IDB.  By
Corollary~\ref{cor:types1}, all tuples realizing a type $\gamma'$ with
$\gamma\preceq\gamma'$, also realize the weaker type $\gamma$. Hence
increasing the size of an IDB relation $P$ by adding new tuples (which
realize a newly added type $\gamma'$) is only possible if in all
present types $\gamma\in\Theta(P,i)$ some atom rank of $\gamma$ is
greater than the corresponding rank in $\gamma'$, i.e.
$\gamma\not\preceq\gamma'$. 

In terms of rank vectors, a type $\gamma$ defines a set
$\mathcal{H}_{\gamma}$ containing the vectors of all types $\gamma'$
that are at least as restrictive as $\gamma$, i.e.  $\gamma\preceq
\gamma'$:
\begin{align*}
  \mathcal{H}_{\gamma} =\condset{(x_1,\dotsc,x_{k_P})}{x_{\ell}\ge
  d_{\ell}^{\gamma}\text{ for }\ell=1,\dotsc,k_P}
\end{align*}

Speaking of rank vectors, $\gamma\preceq \gamma'$ if the rank vector
$\bar{d}^{\gamma}$ is \emph{dominated} by the rank vector
$\bar{d}^{\gamma'}$, i.e. for all $i=1,\dotsc,k_P$,
$d^{\gamma}_i\le d^{\gamma'}_i$. Then
$\mathcal{H}_{\gamma}$ is the set of all types with a rank vector
dominating the rank vector of $\gamma$.

Then creating a sequence of new types added to $\Theta(P,i)$ is
equivalent to the search for a non-dominating sequence of rank
vectors, where we call a (finite or infinite) sequence
$x_1,x_2,\dotsc$ \emph{non-dominating} if for all $i$ and $j$ with
$i<j$, $x_j$ does not dominate $x_i$.

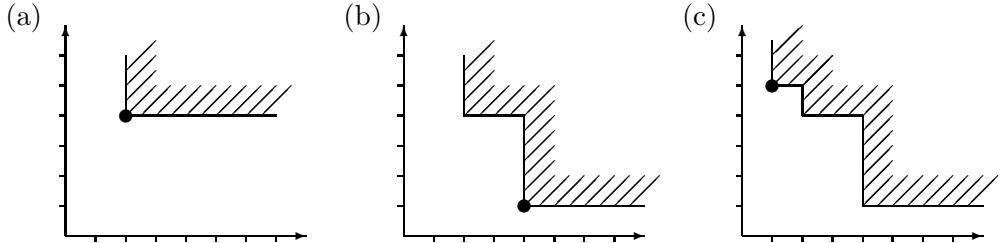
\begin{figure}[htbp]
  \centering  

  \newcommand{\xm}{\circle{0.4}}
  \setlength{\unitlength}{4mm}
  \begin{picture}(9,9)
    \thinlines
    \put(1,1){\vector(1,0){8}}
    \put(1,1){\vector(0,1){7}}
    \multiput(2,1)(1,0){7}{\line(0,-1){0.2}}
    \multiput(1,2)(0,1){6}{\line(-1,0){0.2}}
    
    \put(3,5){\line(0,1){2}}
    \put(3,5){\line(1,0){5}}

    \put(3,5){\circle*{0.4}}

    \multiput(3,5)(.5,0){10}{\line(1,1){1}}
    \multiput(3,5.5)(0,.5){3}{\line(1,1){1}}

    \put(-1,8){(a)}
  \end{picture}
  \qquad
  \begin{picture}(9,9)
    \thinlines
    \put(1,1){\vector(1,0){8}}
    \put(1,1){\vector(0,1){7}}
    \multiput(2,1)(1,0){7}{\line(0,-1){0.2}}
    \multiput(1,2)(0,1){6}{\line(-1,0){0.2}}
    
    \put(3,5){\line(0,1){2}}
    \put(3,5){\line(1,0){2}}
    \multiput(3,5)(.5,0){5}{\line(1,1){1}}
    \multiput(3,5.5)(0,.5){3}{\line(1,1){1}}

    \put(5,2){\line(0,1){3}}
    \put(5,2){\line(1,0){4}}
    \multiput(5,2)(.5,0){8}{\line(1,1){1}}
    \multiput(5,2.5)(0,.5){5}{\line(1,1){1}}

    \put(5,2){\circle*{0.4}}

    \put(-1,8){(b)}
  \end{picture}
  \qquad
  \begin{picture}(9,9)
    \thinlines
    \put(1,1){\vector(1,0){8}}
    \put(1,1){\vector(0,1){7}}
    \multiput(2,1)(1,0){7}{\line(0,-1){0.2}}
    \multiput(1,2)(0,1){6}{\line(-1,0){0.2}}
    
    \put(2,6){\line(0,1){1.5}}
    \put(2,6){\line(1,0){1}}
    \multiput(2,6)(.5,0){3}{\line(1,1){1}}
    \multiput(2,6.5)(0,.5){2}{\line(1,1){1}}    

    \put(3,5){\line(0,1){1}}
    \put(3,5){\line(1,0){2}}
    \multiput(3,5)(.5,0){5}{\line(1,1){1}}
    \multiput(3,5.5)(0,.5){2}{\line(1,1){1}}

    \put(5,2){\line(0,1){3}}
    \put(5,2){\line(1,0){4}}
    \multiput(5,2)(.5,0){8}{\line(1,1){1}}
    \multiput(5,2.5)(0,.5){5}{\line(1,1){1}}

    \put(2,6){\circle*{0.4}}

    \put(-1,8){(c)}
  \end{picture}
  \quad
  
  \mbox{}

  {\small Example of the description of an IDB relation with rank
    vectors of length 2 ($x$ and $y$ coordinate).
    
    Figure (a) shows a description with one rank vector, automatically
    including all types with rank vectors in the hatched area. Figure
    (b) shows the situation after a second rank vector was added,
    automatically including more types. In Figure (c), another vector
    is added.}
  
  \caption{Geometric Representation of a Type Set}
  \label{fig:hypercubes1}
\end{figure}

Figure~\ref{fig:hypercubes1} shows a graphical representation for
$k_P=2$. Figure~\ref{fig:hypercubes1}~(c) shows a case where a new
vector is added containing a coordinate greater than the maximum of
all existing entries. But this growth can only occur in a limited
manner, as we will show. Before, we introduce some notation. 

\begin{defi}
  \label{def:vecnorms}  
  Let $k\in\SetN$ and $\bar{x}=(x_1,\dotsc,x_k)\in \SetN^k$. Then
  $\norm{\bar{x}}:=\max\{x_1,\dotsc,x_k\}$. For $S\subset\SetN^k$, let
  $\norm{S}:=\max_{\bar{x}\in S}\norm{\bar{x}}$. Let
  $s_1,\dotsc,s_{\ell}$ be finite sequences, each sequence consisting
  of tuples of some arity, and let $C=(s_1,\dotsc,s_{\ell})$ be a
  tuple of these sequences. Then
  $\norm{C}:=\max_{i=1}^{\ell}\norm{s_i}$, where the sequences are
  considered as sets.
\end{defi}

To model the rank vectors occurring in the stages of the IDB
relations, we introduce a corresponding sequence concept:

\begin{defi}[\bfseries $c$-Bounded Run]
  \label{def:cboundedrun}
  Let $t\in \SetN$, let $k_1,\dotsc,k_t\in\SetN$ and for $i=1,\dotsc,t$
  let $\bar{x}_i\in\SetN^{k_i}$. Let $c\in\SetN$. Then $X$ is a
  \emph{$c$-bounded run of} $(\bar{x}_1,\dotsc,\bar{x}_t)$, if
  \begin{enumerate}[$\bullet$]
  \item $s^0_1,\dotsc,s^0_t$ are sequences of tuples, where for each $i$,
    $s^0_i$ consists of the tuple $\bar{x}_i$ only.
  \item The stage $X_0$ of $X$ is the tuple $X_0=(s^0_1,\dotsc,s^0_t)$.
  \item Inductively, the $j$-th stage $X_j=(s^j_1,\dotsc,s^j_t)$ of
    $X$ is created from the stage
    $X_{j-1}$ by choosing an
    $\ell\in\{1,\dotsc,t\}$, a\, $\mu_j\in\SetN$, and 
    $\{\bar{x}_1,\dotsc,\bar{x}_{\mu_j}\}\subset\SetN^{k_{\ell}}$ such
    that
    \begin{enumerate}[$-$]
    \item $\mu_j\le (\norm{X_0}\cdot c^{j-1})^{c^2}$
    \item for $n\ne\ell$: $s^j_n=s^{j-1}_n$
    \item $s^j_{\ell}=s^{j-1}_{\ell}\circ
      (\bar{x}_1,\dotsc,\bar{x}_{\mu_j})$\hfill ($\circ$ meaning sequence
      concatenation)
    \item $s^j_{\ell}$ is non-dominating
    \item $\norm{\bar{x}_i}\le c\cdot\norm{X_{j-1}}$ for all
      $i=1,\dotsc,\mu_j$
    \end{enumerate}
  \end{enumerate}
\end{defi}

The condition on $\mu_j$ ensures that the sequence added in each stage
is finite and bounded from above by some function of $\norm{X_0}$, $c$
and $j$, which will be needed for the computation of a uniform bound.
The connection between the setting of datalog programs on orders and
the $c$-bounded runs is given by the following lemma:

\begin{lem}\label{lem:boundsforcboundedrun}
  Let $t$ be the number of nonempty IDB relations of the type-disjoint
  program $\Pi'$ after the initialization sequence $s$ of length $i_s$
  from Lemma~\ref{lem:discinit}. Then for each nonempty IDB relation
  $P$, the set $\Theta(P,i_s)$ contains exactly one rank vector. Let
  $\bar{d}_1,\dotsc,\bar{d}_t$ be these rank vectors. Let
  $m=\max\{m'_R,m'_I,m'_L\}$.

  \begin{enumerate}[\em(1)]
  \item For all $j=1,\dotsc,t$: $\norm{\bar{d}_i}\le (m'_R)^{i_s}\le
    (m'_R)^{n'_R n'_I}$.
  \item For each computation of $\Pi'$ continuing the initialization
    sequence, the rank vectors added during this computation form an
    $m$-bounded run $X$ of $(\bar{d}_1,\dotsc,\bar{d}_t)$.
  \end{enumerate}
\end{lem}

\proof
To prove (1), note that $\norm{\bar{d}_i}\le (m'_R)^{i_s}$ follows from
  Lemma~\ref{lem:discrtypes} and $(m'_R)^{i_s}\le (m'_R)^{n'_R n'_I}$
  follows from Lemma~\ref{lem:discinit}.

  (2) is proved by induction on the steps of the computation. Suppose
  at stage $i$ of the computation of $\Pi$, a rule $\rho$ with IDB $P$
  in its head is applied. Let $\gamma_1',\ldots,\gamma'_{\mu'}$ be the
  types in $\Theta(P,i+1)\setminus \Theta(P,i)$. It may be that some
  of the $\gamma_i'$ dominate $\gamma_j'$ for $j<i$ or
  $\gamma\in\Theta(P,i)$. We omit all these $\gamma_i'$ and obtain a
  sequence $\gamma_1,\ldots,\gamma_\mu$. Adding their rank vectors to
  the run obtained so far, we obtain a non-dominating sequence. The
  $m$-boundedness of the run follows from Remark~\ref{rem:rank}.
\qed

To show a computable uniform bound on $c$-bounded runs, we need two
well known lemmas which we state here without proof:

\begin{lem}[\bfseries K\"onig's Tree Lemma (see, e.g.,
  \cite{hodges97model,diestel00graphen})]
  \label{lem:koenig}  
  Let $T$ be an \emph{infinite} rooted directed tree with finite
  branching (i.e. each vertex has a finite number of children). Then
  $T$ contains an infinite path starting at the root node.
\end{lem}

\begin{lem}[\bfseries Dickson's Lemma (see, e.g.,
  \cite{harwood06weak,dickson13finiteness})]
  \label{lem:dickson}
  All non-dominating sequences of tuples of natural numbers are
  finite. 
\end{lem}

These finiteness (or infiniteness) properties allow us to compute a
bound on the number of stages of $c$-bounded runs:

\begin{lem}
  \label{lem:cboundedruns}
  
  There is a nondecreasing computable function
  $f:\SetN\times\SetN\times\SetN\times\SetN\rightarrow\SetN$ with the
  following property:
  
  For all $m\in\SetN$, $c\in\SetN$, $t\in\SetN$, 
  $r\in\SetN$, $k_1,\dotsc,k_t\in\{1,\dotsc,r\}$ and
  $\bar{x}_i\in\SetN^{k_i}$ with $\norm{\bar{x}_i}\le m$,
  each $c$-bounded run of $(\bar{x}_1,\dotsc,\bar{x}_t)$ has at most
  $f(m,c,t,r)$ stages.
\end{lem}

\proof
  First, we have a look at an arbitrary choice of $m$, $c$, $t$, $r$,
  $k_1,\dotsc,k_t$ and $\bar{x}_i$ (for $i=1,\dotsc,t$).
  
  We create a labeled tree $T$ containing all $c$-bounded runs of
  these values: The root node is labeled with $X_0$. Inductively, for
  each node labeled with a stage $X_i$, we create a child node for
  each stage $X_{i+1}$ created from $X_i$ and label it with the
  corresponding stage.
  
  To create a stage $X_{i+1}$ from $X_i$, we may choose each of the
  $t$ sequences to extend it. Each sequence $s^i_j$ has an arity $k_j$
  and by the last condition of a $c$-bounded run, the element added to
  this sequence may only have coordinates that are at most
  $c\norm{X_i}$. Because of this and since the length of $\mu_{i+1}$
  of the extension of the sequence in stage $i+1$ is bounded from
  above by $(\norm{X_0}\cdot c^{j-1})^{c^2}$, there are only finitely
  many choices for a finite extension of a sequence and hence finitely
  many children for each node in this tree $T$ (each for a different
  extension of some sequence).
  
  A path in $T$ (starting at the root node) corresponds to one
  $c$-bounded run. We now show that each path is finite: Assume, we
  have an infinite path $p$ in $T$. This path $p$ is labeled with the
  stages of a $c$-bounded run $X$. In each stage one of the sequences
  of $X$ is extended by finitely many elements and since there are
  only $t$ sequences there has to be one sequence that is extended in
  infinitely many stages. Each extension of this sequence is
  non-dominating, so we get an infinite non-dominating sequence. But
  by Dickson's Lemma each non-dominating sequence of tuples of natural
  numbers is finite, a contradiction. Hence all paths in $T$ are
  finite.
  
  Hence $T$ has finite branching (only finitely many children to each
  node) and no infinite path. By K\"onig's Lemma $T$ must be finite.
  
  The height of $T$ is the greatest number of stages that can occur in
  a $c$-bounded run of $\bar{x}_1,\dotsc,\bar{x}_k$. Since $T$ is
  finite, we can compute the whole tree and determine its height.
  
  We discuss how to compute the value $f(m,c,t,r)$ for given values
  $m,c,t$ and $r$:
  
  For each fixed choice $k_1,\dotsc,k_t$ of arities, the entries of
  all choices of the corresponding tuples $\bar{x}_1,\dotsc,\bar{x}_t$
  are bounded by $m$ and thus for each tuple there are only finitely
  many choices. By computing the height of the tree (by creating the
  tree) to each choice of tuples one after the other and determining
  the maximum $h(k_1,\dotsc,k_t)$, we have computed a bound on the
  number of stages for the $c$-bounded runs with sequence arities
  $k_1,\dotsc,k_t$.
  
  The parameter $t$ determines the number of sequences in the runs
  considered and the parameter $r$ limits the arities of these
  sequences. The maximum of the values $h(k_1,\dotsc,k_t)$ over all
  possible $r^t$ sequence arity tuples $(k_1,\dotsc,k_t)$ is then the
  maximal number of stages in a $c$-bounded run with $t$ sequences and
  sequence arities at most $r$ and it can be computed by computing
  $h(k_1,\dotsc,k_t)$ for all finitely many choices.
  
  This maximum satisfies the properties of $f(m,c,t,r)$, and that $f$
  is nondecreasing is immediate: Increasing some parameter,
  all runs remain valid, but also longer runs may appear. 
\qed

This function will directly lead to the function $b$ of
Theorem~\ref{theo:ucl}:

\proof \textit{(of Theorem~\ref{theo:ucl})}
  The program $\Pi$ over a linear ordering $\mathcal{A}=(A,<)$ is
  first converted to an equivalent type-disjoint version $\Pi'$ as in
  Lemma~\ref{lem:disjoint}, which also gives the bounds $n'_I\le
  n_I\cdot 3^{m_L^2}$, $n'_R\le 3^{m_L^2\cdot (m_I+1)}\cdot n_R$,
  $m'_R=m_R$ and $m'_L=m_L$ for the parameters of the new program
  $\Pi'$.  Then the initialization sequence $s$ as in
  Lemma~\ref{lem:discinit} is determined, resulting in the first
  $i_s\le n'_I\le n_I\cdot 3^{m_L^2}$ stages.
  
  While the empty relations of $\Pi'$ can be neglected, each nonempty
  relation $P$ of $\Pi'$ has a type description $\Theta(P,i)$ with one
  rank vector each, and by Lemma~\ref{lem:boundsforcboundedrun} these
  rank vectors satisfy the properties of an $m_R$-bounded run $X$. By
  Lemma~\ref{lem:cboundedruns}, $X$ has at most $f((m'_R)^{n'_R
    n'_I},m'_R,n'_I,m'_L)$ stages.
  
  We let $b(n):=f(n^{3^{n4}},n,n\cdot 3^{n^2},n)$ and by the above
  bounds on the parameters and since each program parameter is bounded
  from above by the program length $n$, \linebreak $f((m'_R)^{n'_R
    n'_I},m'_R,n'_I,m'_L)\le f(n^{3^{n4}},n,n\cdot 3^{n^2},n)\le b(n)$
  and the claim follows.
  
  Since $\mathcal{A}$ was chosen as arbitrary linear order, this bound
  also holds for $\ucl(\Pi,\text{LO})$.
\qed

It now follows easily that the datalog tuple problem is decidable on
all linear orders, provided that the orders satisfy the effectivity
conditions described in Section~\ref{sec:problems}, which say that the
elements of a structure are effectively represented, and that the
distance-$d$ type
of a tuple can be computed.

\begin{thm}\label{theo:tupledecidable}
  The datalog tuple problem on linear orders is decidable.
\end{thm}

\begin{proof}
  Using Lemma~\ref{lem:discrtypes} and and Theorem~\ref{theo:ucl}, for
  each IDB we can compute the set of all complete types of tuples that
  are contained in an IDB-relation after the computation of the dalog
  program has been carried out. Then to decide whether a given tuple
  is contained in an IDB relation, we only have to check if it
  satisfies one of these types.
\end{proof}

The uniform closure ordinal of a datalog program can be also be used
to decide the nonemptiness problem. By the EXPTIME-hardness of
nonemptiness, it follows that the uniform closure ordinal of a datalog
program over linear orders has to be at least singly exponential. We
suspect that this lower bound is closer to the actual closure ordinal
than our computable upper bound.

On dense linear orders without endpoints, we can match the singly exponential
lower bound. Let \textit{DLO} denote the class of dense linear orders without
endpoints. 

\begin{thm}\label{theo:ucldense}
  Let $\Pi$ be a datalog program over the class of linear orders and
  $n=|\Pi|$. Then 
  \[
  \ucl(\Pi,\text{DLO})\le 3^{n^2}.
  \]
\end{thm}

\proof
  Observe that on dense linear orders, distance types collapse to
  order types, because for all $a,b$ and for all $d\ge 1$ we have
  $a<b\iff a\le_d b$. So the only types to consider are
  distance-1-types (including equality atoms, when we consider
  complete distance-1-types) and the as type-disjoint version of a
  program as introduced in Lemma~\ref{lem:disjoint} contains all
  complete distance-1-types, it contains all types of interest for
  this case.  After evaluating the initialization sequence as computed
  in Lemma~\ref{lem:discinit}, all complete distance-1-types
  describing the IDB relations are computed and hence no rule can be
  applied after that to add new types.

  Since there at most $3^{n^2}$ different distance-1-types for a
  program of length $n$, the claim follows.\qed 

\section{Relational structures with constants}
\label{sec:DLOconst}

We may also consider datalog programs over linear orders
$\mathcal{A}=(A,<,c_1,\dotsc,c_r)$ with finitely many constants
$c_1,\dotsc,c_r$, each of them being interpreted as a fixed element of
$A$, which may be used in the datalog programs in question.

To solve the datalog nonemptiness or tuple problem for such a program
$\Pi$ with constant symbols, we transform the program to a constant
free version $\Pi'$ by replacing each constant $c_i$ by a fresh
variable and adding rule parts to transfer the values of all constants
to all rules which are used during program execution. Using this
technique, we show:

\begin{thm}\label{theo:constants}
  The datalog tuple problem on linear orders
  $\mathcal{A}=(A,<,c_1,\dotsc,c_r)$ with finitely many constants
  (which may be used by the datalog programs) is decidable.
\end{thm}

\proof We first transform the program $\Pi$ over
$\mathcal{A}=(A,<,c_1,\dotsc,c_r)$ to a program $\Pi'$ over the
structure $\mathcal{A}'=(A,<)$ increasing the arity of each IDB symbol
by $r$, such that for each IDB $P$ of $\Pi$ with arity $s$ and its
corresponding IDB $P'$ of $\Pi'$, and each $\bar{a}=(a_1,\dotsc,a_s)\in
A^s$ the following holds:

\begin{align}
  \label{eq:constrepl}
  \bar{a} \in P^{\Pi,\mathcal{A}}_{\infty} \qquad \Longleftrightarrow
  \qquad
  \bar{c}\bar{a}=(c_1^{\mathcal{A}},\dotsc,c_r^{\mathcal{A}},a_1,\dotsc,a_s)
  \in {(P')}^{\Pi',\mathcal{A}'}_{\infty}
\end{align}

This is established by replacing all occurrences of the constant $c_i$
by a fresh variable $C_i$, for $i=1,\dotsc,r$. To ensure, that all
rule applications share the same values for the constants, we augment
each IDB $P$ of $\Pi$ by the additional variables
$\bar{C}=(C_1,\dotsc,C_r)$ and replace all occurrences of $P(\bar{x})$
by $P'(\bar{C},\bar{x})$ \,---\, in the rule bodies and the rule
heads.  This forces the values of the variables $C_1,\dotsc,C_r$ to be
identical in the body and head of each rule and hence in the whole
pogram.  For example, if
$\phi(c_1,\dotsc,c_r,x_1,\dotsc,x_m,y_1,\dotsc,y_n)$ is the formula
appearing in the body of the rule
\begin{align*}
  P(x_1,\dotsc,x_m) \rd \phi(c_1,\dotsc,c_r,x_1,\dotsc,x_m,y_1,\dotsc,y_n).
\end{align*}
we translate this rule to:
\begin{align*}
  P'(C_1,\dotsc,C_r,x_1,\dotsc,x_m) \rd
  \phi(C_1,\dotsc,C_r,x_1,\dotsc,x_m,y_1,\dotsc,y_n). 
\end{align*}

Now the original program $\Pi$ and the modified version $\Pi'$ satisfy
condition (\ref{eq:constrepl}).

This transformation can be carried out in logarithmic space, since the
number $r$ of constants does only depend on the structure
$\mathcal{A}$, not the input $(\Pi,P,\bar{a})$ of the tuple problem.
The above construction is a logspace reduction from the tuple problem
over linear orders with constants to the tuple problem over linear
orders without constants, mapping the input $(\Pi,P,\bar{a})$ to the
instance $(\Pi',P',\bar{c}\bar{a})$, increasing $m_L$ and $m_R$ by $r$
and the total program size by a linear factor. For this constant free
version of the tuple problem, Theorem~\ref{theo:tupledecidable} shows
us how to solve it, calculating the type sets introduced in
Lemma~\ref{lem:discrtypes}. \qed

We can also use these type sets computed in the above proof for
solving the nonemptiness problem on
$\mathcal{A}=(A,<,c_1,\dotsc,c_r)$:

\begin{cor}
  
  The datalog nonemptiness problem on linear orders
  $\mathcal{A}=(A,<,c_1,\dotsc,c_r)$ with finitely many constants is
  decidable.
\end{cor}

\proof On input $(\Pi,P)$, we first calculate the type sets for the
modified version $\Pi'$, in which the constants have been replaced by
the variables $C_1,\dotsc,C_r$ as above. Then we instantiate the
variables $C_1,\dotsc,C_r$ by the constants of $\mathcal{A}$ in all
types in $\Theta(P',\infty)$ for the IDB $P'$ of $\Pi'$ corresponding
to IDB $P$ of $\Pi$, each $C_i$ by $c^{\mathcal{A}}_i$.
Types which are then no longer satisfiable are deleted from the set.
If and only if there are satisfiable types remaining,
$P^{\Pi,\mathcal{A}}_{\infty}$ is nonempty.  \qed

An intuitive argument, why the upper bound for the datalog
nonemptiness problem on orders with constants is much higher than the
complexity of the case without constants is, that as soon as constants
are involved, the solution process has to consider distances between
constants. A solution may for some two constants $c_i<c_j$ require a
number of elements to be present in $\mathcal{A}$ between $c_i$ and
$c_j$, which is higher than the actual number of elements between
$c_i$ and $c_j$ in $\mathcal{A}$. This case is only handled correctly
by using distance types, order types alone do not suffice. 

However, on dense linear orderings we may do better and match the
EXPTIME lower bound:

\begin{cor}
  The datalog nonemptiness problem on dense linear orders \linebreak
  $\mathcal{A}=(A,<,c_1,\dotsc,c_r)$ with finitely many constants can
  be decided in exponential time.
\end{cor}

\proof
This result is a straightforward combination of the preceeding proof
and Theorem~\ref{theo:ucldense}.
\qed

Note that even though we use the decidability of the datalog tuple problem to
prove the decidability of the datalog nonemptiness problem for linear orders
with constants, we do not need to make any effectivity assumptions on the
linear order $\mathcal A$ (cf. Sec.~\ref{sec:problems}) here. The reason is
that the constants are fixed in advance as part of the structure and thus all
information about them can be hardwired into the algorithm.

\section{Concluding remarks}
\label{sec:conclus}

We studied the complexity of datalog on linear orders. We precisely
determined the complexity of the datalog nonemptiness problem: It is
EXPTIME-complete on all linear orders with at least two elements. We
also obtained a computable uniform upper bound on the closure ordinal
of datalog programs on linear orders. Then best lower bound we know for the
uniform closure ordinals is singly exponential.

The upper bound on the closure ordinals can be used to prove that the
datalog tuple problem is decidable on computable orders leading to the
same complexity bound for the nonemptiness and tuple problems on
orders with constants.  Based on these results, an implementation of
the distance type concept for calculations seems feasible for
applications, e.g. in temporal and spatial reasoning.

In his forthcoming PhD-thesis~\cite{schwa08}, the second author showed
that most of the results obtained here can be extended to colored
linear orders, that is, linear orders with additional unary
predicates, and, at least partially, to colored trees, where trees are
viewed as partial orders.

\section{Acknowledgements}
The authors thank Mark Weyer for proposing an elegant idea leading to
the proof of Theorem~\ref{theo:ucl}.

%
%

\bibliographystyle{plain}
\bibliography{datalogOrders}

\end{document}